\documentclass[floats,floatfix,showpacs,amssymb,prd,superscriptaddress,twocolumn,aps,nofootinbib]{revtex4-1}
\usepackage{graphicx, epsfig, amssymb} 
\usepackage{amsmath, amsfonts}
\usepackage{bm} 
\usepackage[breaklinks]{hyperref}
\usepackage{color}

\def\nn{\nonumber}
\def\be{\begin{equation}}
\def\ee{\end{equation}}
\def\beq{\begin{eqnarray}}
\def\eeq{\end{eqnarray}}
\def\lp{{\ell+1}}
\def\lm{{\ell-1}}
\def\lpp{{\ell+2}}
\def\lmm{{\ell-2}}

\def\ii{{{\rm i}\,}}

\def\cQ{{\cal Q}}

\def\cQ{{\cal Q}}

\begin{document}

\title{Black-Hole Bombs and Photon-Mass Bounds}

\author{Paolo Pani}\email{paolo.pani@ist.utl.pt}
\affiliation{CENTRA, Departamento de F\'{\i}sica, Instituto Superior T\'ecnico, Universidade T\'ecnica de Lisboa - UTL,
Avenida~Rovisco Pais 1, 1049 Lisboa, Portugal.}

\author{Vitor Cardoso}
\affiliation{CENTRA, Departamento de F\'{\i}sica, Instituto Superior T\'ecnico, Universidade T\'ecnica de Lisboa - UTL,
Avenida~Rovisco Pais 1, 1049 Lisboa, Portugal.}
\affiliation{Department of Physics and Astronomy, The University of Mississippi, University, MS 38677, USA.}

\author{Leonardo Gualtieri}
\affiliation{Dipartimento di Fisica, Universit\`a di Roma ``La Sapienza'' \& Sezione, INFN Roma1, P.A. Moro 5, 00185, Roma, Italy.}

\author{Emanuele Berti}
\affiliation{Department of Physics and Astronomy, The University of Mississippi, University, MS 38677, USA.}
\affiliation{California Institute of Technology, Pasadena, CA 91109, USA}

\author{Akihiro Ishibashi}
\affiliation{Theory Center, Institute of Particle and Nuclear Studies,
High Energy Accelerator Research Organization (KEK), Tsukuba, 305-0801, Japan}
\affiliation{Department of Physics, Kinki University, Higashi-Osaka 577-8502, Japan}

\begin{abstract} 
Generic extensions of the standard model predict the existence of
ultralight bosonic degrees of freedom. Several ongoing experiments are
aimed at detecting these particles or constraining their mass
range. Here we show that massive vector fields around rotating black
holes can give rise to a strong superradiant instability which
extracts angular momentum from the hole. The observation of
supermassive spinning black holes imposes limits on this mechanism.
We show that current supermassive black hole spin estimates provide
the tightest upper limits on the mass of the photon ($m_v\lesssim
4\times10^{-20}$~eV according to our most conservative estimate), and
that spin measurements for the largest known supermassive black holes
could further lower this bound to $m_v\lesssim 10^{-22}$~eV. Our
analysis relies on a novel framework to study perturbations of
rotating Kerr black holes in the slow-rotation regime, that we
developed up to second order in rotation, and that can be extended to
other spacetime metrics and other theories.
\end{abstract}


\maketitle

\noindent{\bf{\em Introduction.}}
The properties of matter making up our universe are mostly
unknown. Strong evidence (e.g. from galactic rotation curves and from
gravitational lensing) points to the existence of elusive,
weakly-interacting matter as the most abundant element in the
universe. An interesting possibility is the existence of ultralight
bosonic degrees of freedom, such as those appearing in the ``string
axiverse'' scenario \cite{Arvanitaki:2009fg,Arvanitaki:2010sy}, or of
massive hidden $U(1)$ vector fields, that are also a generic feature
of extensions of the standard model
\cite{Goldhaber:2008xy,Goodsell:2009xc,Jaeckel:2010ni,Camara:2011jg}.

Massive fields around rotating black holes (BHs) can trigger a
superradiant instability, the so-called ``black hole
bomb''~\cite{Press:1972zz}.  This instability is well understood in
the case of massive scalar fields
\cite{Damour:1976kh,Zouros:1979iw,Detweiler:1980uk,Cardoso:2004nk,
  Cardoso:2005vk,Dolan:2007mj,Rosa:2009ei,Cardoso:2011xi}: it requires
the existence of negative energy states in a region around the BH
known as the ergoregion. The instability is regulated by the
dimensionless parameter $M{{{\mu}}}$ (from now on we set $G=c=1$),
where $M$ is the BH mass and $m_s={{{\mu}}}\hbar$ is the scalar field
mass, and it is most effective when $M{{{\mu}}}\sim 1$ and for
maximally spinning BHs. For a solar mass BH and a field of mass
$m_s\sim 1$~eV the parameter $M{{{\mu}}}\sim10^{10}$, and therefore in
many cases of astrophysical interest the instability timescale is
larger than the age of the universe. Superradiant instabilities strong
enough to be observationally relevant ($M{{{\mu}}}\sim1$) can occur
either for light primordial BHs that may have been produced in the
early universe~\cite{1971MNRAS.152...75H,
  1966AZh....43..758Z,1974MNRAS.168..399C} or for ultralight exotic
particles found in some extensions of the standard model
\cite{Arvanitaki:2009fg,Arvanitaki:2010sy}. In the string axiverse
scenario, massive scalar fields with $10^{-33}~{\rm
  eV}<m_s<10^{-18}~{\rm eV}$ could play a key role in cosmological
models. Superradiant instabilities may allow us to probe the existence
of ultralight bosonic fields by producing gaps in the mass-spin BH
Regge spectrum \cite{Arvanitaki:2009fg,Arvanitaki:2010sy}, by
modifying the inspiral dynamics of compact
binaries~\cite{Cardoso:2011xi,Yunes:2011aa, Alsing:2011er} or by
inducing a ``bosenova'', i.e. a collapse of the axion
cloud~\cite{Kodama:2011zc,Yoshino:2012kn,Mocanu:2012fd}.

The curved spacetime dynamics of massive vector fields 
has only been studied in nonrotating backgrounds
\cite{Gal'tsov:1984nb,Herdeiro:2011uu,Rosa:2011my,Konoplya:2005hr}.
While superradiant instabilities are expected to occur also for
massive vector fields, quantitative investigations have been hampered
by our inability to fully understand the (massive vector) Proca
equation
\begin{equation}
\nabla_\sigma F^{\sigma\rho}-{{\mu}}^2 A^\rho=0\,,
\label{proca}
\end{equation}
where $A_\mu$ is the vector potential, $F_{\mu\nu}=\partial_\mu
A_\nu-\partial_\nu A_\mu$ and $m_v={{\mu}}\hbar$ is the mass of the
vector field.  Note that the Lorenz condition $\nabla_\mu A^\mu=0$ is
automatically satisfied and the Proca field $A_\mu$ propagates 3
degrees of freedom~\cite{Rosa:2011my}.  In a nutshell, the problem is
that Eq.~\eqref{proca} does not seem to be separable in the Kerr
background.

\noindent{\bf{\em Framework.}}
The Proca perturbation problem in the Kerr metric becomes tractable if
we work in the slow-rotation approximation. 
Let us focus on the Kerr metric in Boyer-Lindquist coordinates
%
\begin{eqnarray}
 ds_{\rm Kerr}^2&=&-\left(1-\frac{2Mr}{\Sigma}\right)dt^2+\frac{\Sigma}{\Delta}dr^2-\frac{4rM^2}{\Sigma}\tilde{a}\sin^2\vartheta d\varphi dt   \nonumber\\
&+&{\Sigma}d\vartheta^2+
\left[(r^2+M^2\tilde{a}^2)\sin^2\vartheta +\frac{2rM^3}{\Sigma}\tilde{a}^2\sin^4\vartheta \right]d\varphi^2\,, \nonumber
\end{eqnarray}
where $\Sigma=r^2+M^2\tilde{a}^2\cos^2\vartheta$, $\Delta=(r-r_+)(r-r_-)$, $r_\pm=M(1\pm\sqrt{1-\tilde a^2})$ and $M$
and $J=M^2\tilde a$ are the mass and the angular momentum of the BH,
respectively. In what follows, we shall expand the metric and all
other quantities of interest to second order in $\tilde{a}$.
The procedure to separate the linearized perturbation equations to
first order in $\tilde{a}$ was first proposed by Kojima in the context
of stellar
perturbations~\cite{Kojima:1992ie,1993ApJ...414..247K,1993PThPh..90..977K},
but it can be generalized to \emph{any} order in $\tilde{a}$ and to
generic (scalar, vector, tensor, etc.) perturbations of stationary and
axisymmetric spacetimes.  The details of the procedure will appear
elsewhere~\cite{elsewhere}; here we only present the main results.

In the slow-rotation limit the perturbation equations, expanded in
spherical harmonics and Fourier transformed in time, yield a coupled
system of ODEs. In the case of a spherically symmetric background,
perturbations with different harmonic indices $(\ell,\,m)$, as well as
perturbations with opposite parity, are decoupled. In a rotating,
axially symmetric background, perturbations with different values of
the azimuthal number $m$ are still decoupled, but perturbations with
different values of $\ell$ are not.  However, in the limit of slow
rotation there is a Laporte-like ``selection
rule''~\cite{ChandraFerrari91}: at first order in $\tilde{a}$,
perturbations with a given value of $\ell$ are only coupled with those
with $\ell\pm1$ and \emph{opposite} parity, similarly to the case of
rotating stars. At second order, perturbations with a given value of
$\ell$ are also coupled with those with $\ell\pm2$ and \emph{same} parity,
and so on.

In general, the perturbation equations can always be written in the
form~\cite{elsewhere}
\begin{eqnarray}
0&=&{\cal A}_{\ell}+\tilde a m \bar{\cal A}_{{\ell}}+\tilde{a}^2 \hat{{\cal A}}_\ell\nn\\
&+&\tilde a ({\cal Q}_{{\ell}}\tilde{\cal P}_{\ell-1}+{\cal Q}_{\ell+1}\tilde{\cal P}_{\ell+1})\nn\\
&+&\tilde{a}^2 \left[\cQ_\lm \cQ_\ell \breve{{\cal A}}_\lmm + \cQ_\lpp \cQ_\lp \breve{{\cal A}}_\lpp \right]+{\cal O}(\tilde{a}^3)\,,\label{epF1c}\\
0&=&{\cal P}_{\ell}+\tilde a m \bar{\cal P}_{{\ell}}+\tilde{a}^2 \hat{{\cal P}}_\ell\nn\\
&+&\tilde a ({\cal Q}_{{\ell}}\tilde{\cal A}_{\ell-1}+{\cal Q}_{\ell+1}\tilde{\cal A}_{\ell+1})\nn\\
&+&\tilde{a}^2 \left[\cQ_\lm \cQ_\ell \breve{{\cal P}}_\lmm + \cQ_\lpp \cQ_\lp \breve{{\cal P}}_\lpp \right]+{\cal O}(\tilde{a}^3)\,,\label{epF2c}
\end{eqnarray}
where ${\cal Q}_\ell=\sqrt{\frac{\ell^2-m^2}{4\ell^2-1}}$ and the
coefficients ${\cal A}_\ell$ and ${\cal P}_\ell$ (with various
superscripts) are \emph{linear} combinations of axial and polar
perturbation variables, respectively.

The general method can be specialized to the Proca
equation~\eqref{proca}. We expand the vector potential
as~\cite{Rosa:2011my}
\begin{eqnarray}\label{harmonic_exp}
\delta A_{\mu}(t,r,\vartheta,\varphi)&=&\sum_{\ell,m}\left[
 \begin{array}{c} 0 \\ 0\\
u_{(4)}^\ell \mathbf{S}_b^{\ell}/\Lambda\\
 \end{array}\right]+\left[ \begin{array}{c}u_{(1)}^\ell Y^{\ell}/r\\u_{(2)}^\ell Y^{\ell}/(r f) \\
 u_{(3)}^\ell \mathbf{Y}_b^{\ell}/\Lambda\\ \end{array}\right] \,,\nn
\end{eqnarray}
where $b\equiv(\vartheta,\varphi)$, $\Lambda=\ell(\ell+1)$, $f=\Delta/(r^2+\tilde{a}^2M^2)$, $Y^\ell$ are scalar spherical harmonics, $\mathbf{Y}^\ell_b\equiv(Y_\vartheta^\ell,Y_\varphi^\ell)$ and $\mathbf{S}_b^\ell\equiv(S_\vartheta^\ell,S_\varphi^\ell)$ are vector spherical
harmonics~\cite{elsewhere}, whereas $u_{(i)}^{\ell}=u_{(i)}^{\ell}(r,t)$ $(i=1,2,3)$ and
$u_{(4)}^{\ell}=u_{(4)}^{\ell}(r,t)$ are polar and axial perturbations, respectively.
Separating the angular variables, we find that Proca perturbations in
the slow-rotation limit, up to second order, are described by two sets
of equations~\cite{elsewhere}:
\begin{eqnarray}
 \mathbf{{\cal D}_A}\mathbf{\Psi_A}^\ell+\mathbf{V_A}\mathbf{\Psi_A}^\ell&=&0\,,\label{systA}\\
 \mathbf{{\cal D}_P}\mathbf{\Psi_P}^\ell+\mathbf{V_P}\mathbf{\Psi_P}^\ell&=&0\,,\label{systP}
\end{eqnarray}
where $\mathbf{{\cal D}_{A,P}}$ are second order differential operators,
$\mathbf{V_{A,P}}$ are matrices,
$\mathbf{\Psi_A}^\ell=(u_{(4)}^\ell,u_{(2)}^{\ell\pm1},u_{(3)}^{\ell\pm
  1},u_{(4)}^{\ell\pm2})$ and
$\mathbf{\Psi_P}^\ell=(u_{(2)}^\ell,u_{(3)}^\ell,u_{(4)}^{\ell\pm
  1},u_{(2)}^{\ell\pm 2},u_{(3)}^{\ell\pm 2})$. The function
$u_{(1)}^{\ell}$ can be obtained from the Lorenz condition once the
three dynamical degrees of freedom are known~\cite{elsewhere}. When
$\tilde a=0$, the equations above reduce to Proca perturbations of a
Schwarzschild BH~\cite{Rosa:2011my}. However, rotation introduces
mixing between perturbations of different parity and different
multipolar indices.

\noindent{\bf{\em Numerical Results.}}
Once suitable boundary conditions and a time dependence of the form
$e^{-\ii\omega t}$ are imposed, Eqs.~\eqref{systA}--\eqref{systP} form
an eigenvalue problem for the complex frequency $\omega=\omega_R+\ii
\omega_I$. Physically motivated boundary conditions correspond to
either quasinormal modes (perturbations having ingoing-wave conditions
at the horizon and outgoing-wave conditions at
infinity~\cite{Berti:2009kk}) or bound states (perturbations that are
spatially localized within the vicinity of the BH and decay
exponentially at infinity). Here we focus on bound modes. By analogy
with the scalar field case we would expect these modes to become
superradiantly unstable for
$\omega_R<m\Omega_H$~\cite{Detweiler:1980uk}, where $\Omega_H=\tilde a
/(2 r_+)$. Our analysis shows, for the first time, that massive vector
fields do indeed become unstable in this regime.

The bound state modes of the system~\eqref{systA}--\eqref{systP} can
be found by standard numerical methods~\cite{elsewhere}.  
When $m>0$ we find that, within numerical errors, the imaginary part
of the modes has a zero crossing when
\begin{equation}
 \omega_R=m\Omega_H\sim m\frac{\tilde a}{4 M}+{\cal O}(\tilde a^3)\,,
\label{superradiance}
\end{equation}
which corresponds to the onset of the superradiant regime.  When
$\omega_R<m \Omega_H$ the modes are unstable and $\tau=\omega_I^{-1}$
is the instability growth timescale.  Note that, although the
superradiant condition~\eqref{superradiance} appears as a first order
effect, in fact $\omega M\sim{\tilde a}$ at the onset of
superradiance.  The field equations contain terms proportional to
$\tilde{a}\omega$ and to $\omega^2$, so a second-order expansion is
needed for a self-consistent study of the unstable
regime~\cite{elsewhere}.

When $M{{\mu}}\lesssim 0.1$ our data for the fundamental modes are
consistent with a hydrogenic spectrum, $\omega_R\sim{{\mu}}$ and
\begin{equation}
M\omega_I\sim \gamma_{S\ell}\left(\tilde a m-2 r_+{{\mu}} \right) (M{{\mu}})^{4\ell+5+2S}\,,
\label{fit_wI}
\end{equation}
where $\gamma_{S\ell}$ is a coefficient that depends on $\ell$ and on
the ``polarization'' index $S$, with $S=0$ for axial modes and
$S=\pm1$ for two classes of polar modes~\cite{Rosa:2011my}.  
In the axial case the numerical results are also supported by an
analytical formula, that can be found by applying Starobinski's method
of matching asymptotics~\cite{Starobinsky,Detweiler:1980uk} to
Eq.~\eqref{systA} at first order. A detailed
calculation~\cite{elsewhere} yields precisely Eq.~\eqref{fit_wI} with
$S=0$ and $\gamma_{01}=1/12$. This is consistent with our numerical
data, that yield $\gamma_{01}\approx0.09\pm0.03$ (here and in the
following, numerical errors are estimated by comparing the results at
first and second order and by taking the maximum deviation between the
fit and the data).  The instability timescale in the axial Proca case
is four times shorter than in the scalar case. Similar results can be
derived also for axial modes with $\ell>1$ and for the overtones.

To our knowledge, this is the first estimate of the instability
timescale of massive vector fields around spinning BHs.  Although
Eq.~\eqref{fit_wI} is strictly valid only when $\tilde a\ll1$ and
$M{{\mu}}\ll1$, in the case of massive scalar fields it provides
estimates in good agreement with exact
results~\cite{Cardoso:2005vk,Dolan:2007mj} up to
$\tilde{a}\lesssim0.99$: for example, Eq.~\eqref{fit_wI} overestimates
the exact result by only $3\%$ when $\tilde{a}=0.7$, and by less than
$70\%$ when $\tilde{a}=0.99$~\cite{elsewhere}.

In the massive vector case, the good agreement between the first- and
second-order calculations suggests that the slow-rotation expansion
can be trusted even for moderately large spins,
$\tilde{a}\lesssim0.7$. Therefore it is reasonable to expect that
extrapolations of Eq.~\eqref{fit_wI} from the slow-rotation limit
should at least provide the correct order of magnitude (and possibly a
reliable quantitative estimate) of the instability timescale far from
extremality. If we extrapolate Eq.~\eqref{fit_wI} to $\tilde{a}\to1$,
we expect to overestimate the instability by about one order of
magnitude~\cite{elsewhere}.

As we shall discuss below, astrophysical bounds on vector field masses
(scaling as $\gamma_{S\ell}^{-1/(4\ell+5+2S)}$~\cite{elsewhere})
depend very mildly on uncertainties in the $\gamma_{S\ell}$
coefficients. However they are sensitive to the scaling with $\mu$ in
the $M\mu\ll1$ regime, and it is crucial to obtain reliable results in
this limit. Unfortunately the calculation of unstable modes when
$M\mu\lesssim0.02$ is challenging due to numerical inaccuracies, but
the consistency between our numerical data in the axial case and the
analytical formula (which is valid when $M\mu\ll1$) is reassuring. The
equations governing polar modes are much more complex, and we could
not find analytical results supporting the fit~\eqref{fit_wI} when
$S\neq0$ (cf.~\cite{Rosa:2011my}). Our data for $S=-1$ polar modes at
the onset of the $\ell=m=1$ instability are consistent with
Eq.~\eqref{fit_wI} with $\gamma_{-11}\approx20\pm10$, but some
experimentation showed that different fitting functions can provide
even better fits. Due to these uncertainties in the fit~\eqref{fit_wI}
for the polar case, in the following we shall discuss the consequences
of our results mainly for axial modes, for which we derived the
instability timescale also \emph{analytically}~\cite{elsewhere}. This
choice is very conservative because, according to Eq.~\eqref{fit_wI},
polar modes with $S=-1$ exhibit the strongest instability. Indeed, for
fixed values of $\tilde{a}$ and $\mu$ the instability of polar modes
with $S=-1$ is typically two or three orders of magnitude stronger
than in the axial case~\cite{elsewhere}.

\noindent{\bf{\em Astrophysical bounds on the photon mass.}}
Our results, together with reliable supermassive BH spin measurements,
can be used to impose stringent constraints on the allowed mass range
of massive vector fields. These bounds follow from the requirement
that astrophysical spinning BHs should be stable, in the sense that
the instability timescale $\tau$ should be larger than some
observational threshold. For isolated BHs we can take the
observational threshold to be the age of the Universe, $\tau_{\rm
  Hubble}=1.38\times 10^{10}$~yr. However, for supermassive BHs we may
worry about possible spin growth due to mergers with other BHs and/or
accretion. The most likely mechanism to produce fast-spinning BHs is
prolonged accretion \cite{Berti:2008af}. Therefore, a conservative
assumption to estimate the astrophysical consequences of the
instability is to compare the superradiance timescale to the (minimum)
timescale over which accretion could spin up the BH.  Thin-disk
accretion can increase the BH spin from $\tilde a=0$ to $\tilde
a\simeq1$ with a corresponding mass increase by a factor
$\sqrt{6}$~\cite{Bardeen:1970zz}. If we assume that
mass growth occurs via accretion at the Eddington limit, so that the
BH mass grows exponentially with $e$-folding time given by the
Salpeter timescale $\tau_{\rm Salpeter}=4.5\times 10^7$~yr, then the
minimum timescale for the BH spin to grow via thin-disk accretion
is comparable to $\tau_{\rm Salpeter}$.

\begin{figure}[thb]
\begin{center}
 \epsfig{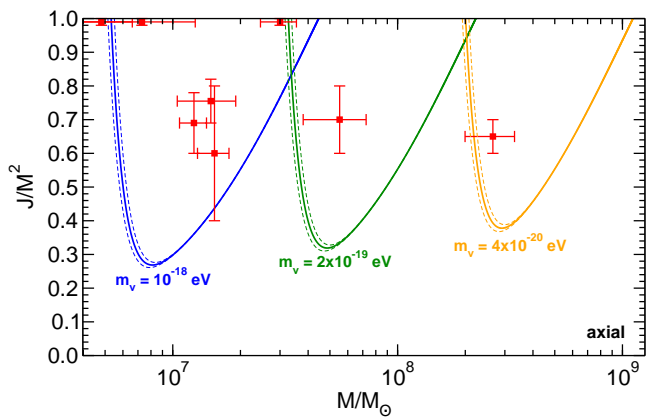}
\caption{Contour plots in the BH Regge plane~\cite{Arvanitaki:2010sy}
  corresponding to an instability timescale shorter than $\tau_{\rm
    Salpeter}$ for different values of the vector field mass
  $m_v={{\mu}}\hbar$ and for axial modes with $\ell=m=1$. Dashed lines
  bracket our estimated numerical errors,
  $\gamma_{01}\approx0.09\pm0.03$ in Eq.~\eqref{fit_wI}.  The
  experimental points (with error bars) refer to the supermassive BHs
  listed in Table~2 of~\cite{Brenneman:2011wz} and the rightmost point
  corresponds to the supermassive BH in
  Fairall~9~\cite{Schmoll:2009gq}. Supermassive BHs lying above each
  of these curves would be unstable on an observable timescale, and
  therefore each point rules out a range of Proca field
  masses.\label{fig:bound2}
}
\end{center}
\end{figure}

Brenneman et al.~\cite{Brenneman:2011wz} have recently presented a
list of eight supermassive BH spin estimates.  In order to quantify
the dependence of Proca field mass bounds on the mass and spin of
supermassive BHs, in Fig.~\ref{fig:bound2} we show exclusion regions
in the ``BH Regge plane'' (cf. Fig.~3 of~\cite{Arvanitaki:2010sy}). To
be more specific, we plot contours corresponding to an instability
timescale of the order of the Salpeter time for three different masses
of the Proca field ($m_v=10^{-18}$~eV, $2\times10^{-19}$~eV,
$4\times10^{-20}$~eV) and for axial modes ($S=0$). The plot shows
that observations of supermassive BHs with $10^6M_\odot\lesssim
M\lesssim 10^9M_\odot$ and $\tilde a\gtrsim 0.3$ would exclude a wide
range of vector field masses.
\begin{figure}[thb]
\begin{center}
 \epsfig{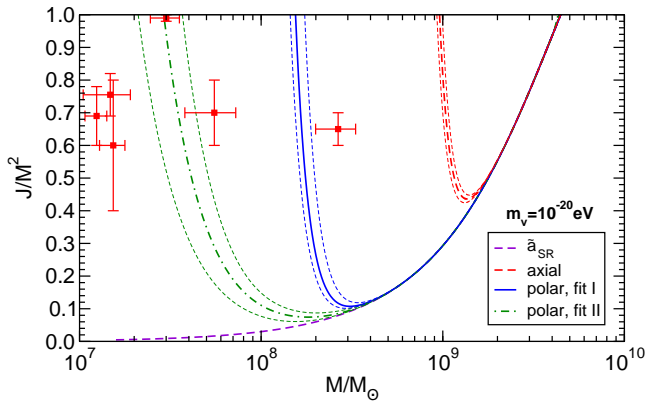}
\caption{Comparison between axial and polar $\ell=m=1$ instability
  windows for $m_v=10^{-20}{\rm eV}$. The right boundary of the
  instability window does not depend on uncertainties in the fits and
  it is given by $J/M^2\equiv \tilde{a}\sim\tilde a_{\rm SR}\sim{4
    M\mu}/{m}+{\cal O}(\mu^3)$, corresponding to the superradiance
  threshold [Eq.~\eqref{superradiance}] when $\omega_R\sim\mu$.  For
  polar modes we show two different fitting functions. Fit~I
  corresponds to Eq.~\eqref{fit_wI} with $\gamma_{-11}\approx20\pm10$,
  i.e, $M\omega_{I}=20({\tilde a}-2r_+\mu)(M\mu)^{7}$. Fit~II is given
  by $M\omega_I\sim\left(\tilde a -\tilde a_{\rm SR}\right)
  \left[\eta_0(M\mu)^{\kappa_0}+\eta_1\tilde{a}
    (M\mu)^{\kappa_1}\right]$ with $\eta_0\approx-6.5\pm2$,
  $\eta_1\approx 2.1\pm1$, $\kappa_0\approx 6.0\pm0.1$,
  $\kappa_1\approx 5.0\pm0.3$. While fit~I is physically more
  appealing \cite{Rosa:2011my}, fit~II does a better job at
  reproducing our numerical data in the whole instability region.
\label{fig:comparison}
}
\end{center}
\end{figure}

In Fig.~\ref{fig:comparison} we compare the axial and polar
instability windows for $m_v=10^{-20}{\rm eV}$, including our
estimated errors. In the polar case we use two different fitting
functions, to bracket uncertainties. The polar instability window
clearly depends on the chosen fitting function, but some general
conclusion can be drawn: (i)~axial bounds on $m_v$ are typically less
stringent than polar bounds by up to one order of magnitude or more;
(ii)~different fitting functions translate into bounds on $m_v$ that
differ by a factor of a few; (iii)~the instability windows in the
polar case extend down to $\tilde{a}\sim0.1$. Thus, essentially
\emph{any} supermassive BH spin measurement would exclude a
considerable range of $m_v$; (iv)~from Eq.~\eqref{superradiance} it is
straighforward to obtain a conservative \emph{upper} bound for the
boson mass that can be excluded by a given observation:
\begin{equation}
 m_v\lesssim m_v^{(c)}=\frac{m\tilde{a}\hbar}{4M}\sim 3.34\times 10^{-19} \,m{\tilde a} \frac{10^8 M_\odot}{M}\, {\rm eV}\,.
\end{equation}
This bound is valid independently of
uncertainties in the fitting functions.

Thus, {\em existing} measurements of supermassive BH spins rule
out vector field masses in the whole range $10^{-20}$~eV~$\lesssim
m_v\lesssim 10^{-17}$~eV. The best bound comes from Fairall~9
\cite{Schmoll:2009gq}, for which the axial instability implies a
conservative bound (including measurement and numerical errors)
$m_v\lesssim 4\times10^{-20}$~eV when we compare the instability
timescale to the Salpeter time, and $m_v\lesssim 2\times 10^{-20}$~eV
if we do not consider accretion. This result is of great significance,
since it is two orders of magnitude more stringent than the current
best bound on the photon mass,
$m_\gamma<10^{-18}$~eV~\cite{PDG}.
If the largest known supermassive BHs with $M\simeq 2\times 10^{10}
M_\odot$ \cite{2011Natur.480..215M,2012arXiv1203.1620M} were confirmed
to have nonzero spin, we could get bounds as low as $m_v\lesssim
10^{-22}$~eV.

\noindent{\bf{\em Nonlinear effects, other couplings.}}
An important ingredient that was not taken into account in our study
is the nonlinear evolution of the instability, that can modify the
background geometry. Photon self-interactions are very weak, being
suppressed by the mass of the electron. Therefore, it is quite likely
that the outcome of the instability will be a slow and gradual
drainage of the hole's rotational energy.
Another important issue is whether the coupling of accreting matter to
massive bosons can quench the instability. In principle massive
photons (unlike hidden $U(1)$ fields, for which the interaction with
matter is very small) can couple strongly to matter. However it is
unlikely that this will significantly affect the superradiant
instability discussed here, for two reasons: (i)~the unstable modes
are large-scale coherent modes whose Compton wavelength is of order
the BH size or larger, and accretion disks are typically
charge-neutral over these lengthscales, so any possible coupling with
ordinary neutral matter is incoherent and most likely inefficient;
(ii)~accretion disks are often localized in the equatorial plane, and
therefore they can affect at most some (but not all) unstable modes.
The investigation of the superradiant instability in the presence of
matter requires further work, but these arguments suggest that
estimates in vacuum should be robust. Spin measurements for slowly
accreting BHs (such as the BH at the center of our own galaxy) are
presumably the most reliable, being rather insensitive to details of
the interaction of vector fields with matter.

\noindent{\bf{\em Conclusions.}}
The results discussed here show that BHs offer the exciting
possibility to constrain particle physics and to set stringent upper
bounds on the mass of bosonic fields. Our method can be generalized to
other fields and applied to other backgrounds, such as Kerr-Newman BHs
or higher-dimensional BHs in TeV-gravity scenarios. Numerical
simulations of either linearized or nonlinear perturbations of the
Kerr geometry are necessary. In numerical studies of superradiant
instabilities a linear perturbation analysis proved to be very useful,
e.g. for choosing suitable initial
data~\cite{Yoshino:2012kn}. Numerical methods may give us a better
understanding of the maximum instability rate, of the nonlinear
evolution and of the end state of the instability. These are clearly
important topics: for example, a more accurate quantitative analysis
of the polar sector could improve the bounds on the photon mass by by up to one order of magnitude or more.

\noindent{\bf{\em Acknowledgments.}}
We wish to thank the Axiverse Project members (especially Hideo
Kodama) and Antonino Flachi for valuable discussions, and the referees
for their useful suggestions.  This work was supported by the
DyBHo--256667 ERC Starting Grant, the NRHEP--295189
FP7--PEOPLE--2011--IRSES Grant, and by FCT - Portugal through PTDC
projects FIS/098025/2008, FIS/098032/2008, CTE-ST/098034/2008,
CERN/FP/123593/2011. A.I. was supported by JSPS Grant-in-Aid for
Scientific Research Fund No. 22540299 and No. 22244030.  E.B. was
supported by NSF Grant No. PHY-0900735 and NSF CAREER Grant
No. PHY-1055103. P.P. acknowledges financial support provided by the
European Community through the Intra-European Marie Curie contract
aStronGR-2011-298297 and the kind hospitality of the University of
Rome ``Sapienza'' and of the International School for Advanced Studies
(SISSA) in Trieste. V.C. thanks the Yukawa Institute for Theoretical
Physics at Kyoto University for hospitality during the YITP-T-11-08
workshop on ``Recent advances in numerical and analytical methods for
black hole dynamics''.

\bibliography{proca}
\end{document}